\documentclass[amsmath,amssymb,floats,twocolumn]{jpsj3}
\usepackage{times}
\usepackage{epsfig}
\usepackage{graphicx}
\usepackage{pstricks}
\usepackage{dcolumn}		
\usepackage{bm}			
\usepackage{units}		
\usepackage{color}
\usepackage{multirow}

\begin{document}

\title{Intertwined states at finite temperatures in the Hubbard model} 
\author{E. W. Huang$^{1}$, W. O. Wang$^{2,3}$, J. K. Ding$^{2,3}$, T. Liu$^{2,4}$, F. Liu$^{2,5}$, X.-X. Huang$^{2,3}$, B. Moritz$^{2}$, T. P. Devereaux$^{2,6,7}$}
\inst{$^{1}$ Department of Physics and Institute of Condensed Matter Theory, University of Illinois at Urbana-Champaign, Urbana, Illinois 61801, USA \\
$^2$Stanford Institute for Materials and Energy Sciences, SLAC National Accelerator Laboratory, 2575 Sand Hill Road, Menlo Park, CA 94025, USA \\
$^{3}$ Department of Applied Physics, Stanford University, Stanford, California 94305, USA \\
$^{4}$ Department of Chemistry, Stanford University, Stanford, California 94305, USA \\
$^{5}$ Department of Physics, Stanford University, Stanford, California 94305, USA \\
$^6$Department of Materials Science and Engineering, Stanford University, CA 94305, USA \\
$^7$Geballe Laboratory for Advanced Materials, Stanford University, CA 94305, USA \\}
\date{\today}

\abst{Significant advances in numerical techniques have enabled recent breakthroughs in the study of various properties of the Hubbard model - a seemingly simple, yet complex model of correlated electrons that has been a focus of study for more than half a century.  In particular, it captures the essence of strong correlations, and is believed to possess various emergent, low energy states and collective excitations characteristic of cuprate high-temperature superconducting materials.  While a thorough review of all activity is not possible here, we have focused the discussion on our recent work using unbiased, numerically exact, ``brute force", finite temperature quantum Monte Carlo methods.  Our various studies reveal a rich variety of quantum liquid crystal phases, and complementary transport properties, which answer some questions, but certainly raise others concerning ``strange metal" behavior and the ultimate fate of quasiparticles in the Hubbard model.}
\maketitle

\section{Introduction}
Several of the most exciting problems in the area of ``quantum materials", including spin liquid behavior and superconductivity, are not well captured by standard approximations, {\it e.g.} density functional theory (DFT) and related methods, as the full many-body wavefunction lies very far away from a simple description in terms of single-electron states. For example, in spin liquids the emergence of fractional quasiparticles, ground state degeneracy, and other features of topological order are not apparent in the single-particle spectral or Green's function. This places a premium on methods that can capture the full Hilbert space of a quantum material and yet scale more efficiently than ``brute force" diagonalization of the Hamiltonian.  

The past decade has seen tremendous advances in numerical methods applied to studying properties of the Hubbard model - the simplest model Hamiltonian for correlated systems that has resisted exact solutions in two dimensions for almost six decades\cite{Hubbard1963}. It is now commonplace to find discussions of the Hubbard model in many textbooks\cite{Maekawa}.  However, in the last decade algorithms, both new and old, have provided a wealth of new information that has deepened our understanding of emergence in this simple Hamiltonian. Unbiased and exact methods, such as zero-temperature, density matrix renormalization group (DMRG)\cite{White1992,Stoudenmire2012} and finite-temperature, determinant quantum Monte Carlo (DQMC)\cite{Blankenbecler1981,White1989,santos2003introduction,Hirsch1985}, have provided information that complements that from other methods based on matrix product states and tensor networks\cite{schollwock2011density}.  This does not even begin to touch on other various variational methods and embedding techniques, such as dynamical mean-field methods\cite{RevModPhys.68.13}, all brought to bear on the Hubbard model defined on different lattices and in various dimensions. 

The simple Fermi-Hubbard model on a 2D square lattice captures the essence of strong correlations and has been believed to harbor various emergent, low energy states and collective excitations that are characteristic of the cuprate high-temperature superconducting materials. However, questions remain about whether such a seemingly simple model can lead to the myriad phases seen across the cuprates. Recently, some questions have been answered more definitively using classical computers about whether the model possesses superconductivity and charge (spin) density-wave ordering on 2D small clusters or quasi 1D ladders\cite{Kung2016,Huang2018,YY}; but the model's behavior approaching the thermodynamic limit still remains something of an open question.

Admittedly, a short review article can in no way cover the complexity of proposed ``solutions" of the Hubbard model. Here, we modestly reduce the scope and discuss only our recent work on ``brute force" methods that make use of DQMC. We refer to the reader to the many review articles available for a much broader perspective\cite{santos2003introduction,arovas2021hubbard}. In this article we will present a review of recent results from DQMC simulations of the Hubbard model, exploring various orders and dynamical quantities, to assess in which sense the Hubbard model compares favorably with the phase diagram of the cuprates, and importantly, to shed light on the role and importance of various terms in the Hamiltonian. The outline of the paper is as follows: Section 2 presents a short introduction to the DQMC method, followed in Section 3 with a short review of the nature of magnetism in the doped Hubbard model, utilizing spectroscopy to identify how spin-stripe correlations emerge at temperature scales below the superexchange $J/k_B$; Section 4 summarizes recent results on pair field susceptibilities to search for superconducting and pair density-wave order, whereas Section 5 examines transport properties in this so called ``strange metal" phase at these temperatures. Finally, Section 6 presents a brief outlook for future studies.

\section{A very brief review of the determinant quantum Monte Carlo method}

DQMC exploits the equivalence of an interacting problem to an ensemble of non-interacting problems with auxiliary fields coupled to density \cite{Blankenbecler1981,White1989,santos2003introduction,Hirsch1985}. If a non-negative weight $w_s$ can be associated with each configuration $s$ of auxiliary fields, the Metropolis algorithm can then be used to simulate the interacting problem by sampling over field configurations. 

With sufficient sampling, the Monte Carlo sample mean
\begin{equation}
\langle A \rangle_{\text{MC}} =  \overline{\langle A \rangle_s}
\end{equation}
is considered an unbiased estimator of the thermal expectation value of operator $A$. As such, DQMC allows us to study a wide range of quantities of interest in the Hubbard model at finite temperature. Because DQMC only attempts to obtain expectation values of operators instead of exact wavefunctions and eigenvalues, it allows us to study larger system sizes than direct diagonalization of the Hubbard Hamiltonian.

\begin{figure}
\includegraphics[width=\columnwidth]{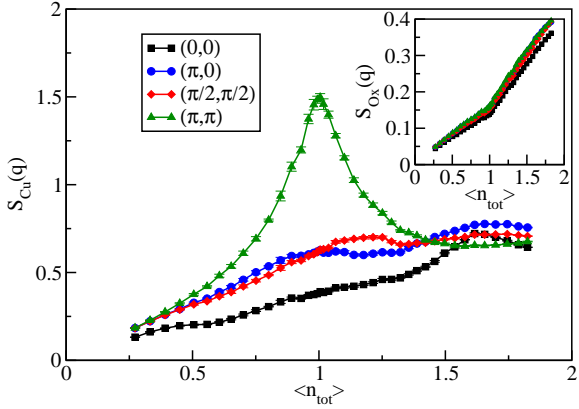}
\caption{The spin-spin correlation function of the three-band Hubbard model is plotted versus filling for four different possible ordering vectors on the copper and oxygen (inset) orbitals. Parameters used here are $N=4 \times 4$ CuO$_2$ unit cells, $U_{dd}=8.5$ eV, $U_{pp}=4.1$ eV, $t_{pd} = 1.13$ eV, $t_{pp} = 0.49$ eV, $\Delta_{pd} = 3.24$ eV, and $\beta=10$ eV$^{-1}$.\cite{Kung2016}}
\label{f1}
\end{figure}

The single largest caveat of working with DQMC is the fermion sign problem \cite{loh1990sign}, an unavoidable reality of simulating interacting quantum many-body systems on a classical computer. In DQMC, the configuration weights must undergo a re-weighting procedure in order to ensure that they are non-negative. This leads to exponential time scaling as a function of simulation size, both
spatial and temporal ({\it i.e.} inverse temperature), and ultimately limits the accessible parameter regime.
A typical Monte Carlo ``run", at least for the results presented here, consists of first, several thousand ``warm-up" Monte Carlo sweeps, that ensure equilibriated sampling for a given set of parameters, and second, between several hundred thousand and a few million ``measurement" Monte Carlo sweeps, depending on temperature and quantities of interest. 

\section{Spin stripe correlations in the Hubbard model}

\begin{figure}
\centering
\includegraphics[width=0.8\columnwidth]{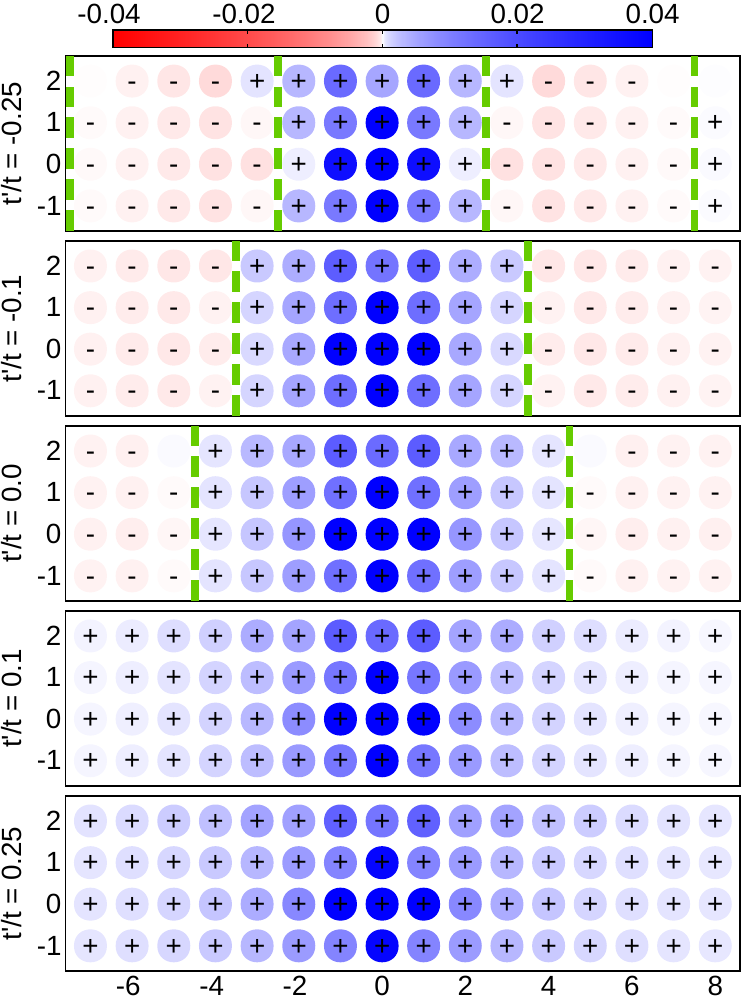}
\caption{Staggered spin correlation functions $\tilde S(i,j)$ of the single-band Hubbard model from DQMC at $p=0.125$ hole doping, $U/t=6$, and $T/t=0.22$ for various values of $t'$.\cite{Huang2018}}
\label{f3}
\end{figure}

Charge and spin stripe correlations appear to be universal in cuprate
superconductors over a broad range of their phase diagram\cite{Tranquada1995,Comin2016,Tranquada1995,Tranquada1994,Zaliznyak2000,Ramirez1996}. It remains an open question whether there is a direct link
between fluctuating stripes and the mechanism
of high-$T_c$ superconductivity \cite{Emery1997,Kivelson1998,Zaanen2001,Kivelson2003,Cvetkovic2007}. The 
hourglass-shaped magnetic excitation spectrum
\cite{Tranquada2007,Vojta2009} found both in compounds that exhibit static stripe order \cite{Tranquada2004} and in those that do
not \cite{Hayden2004,Xu2009} finds a natural explanation in the concept of fluctuating stripes \cite{Kivelson2003,Zaanen1996}. It is important to draw a distinction between incommensurate magnetic fluctuations based on itinerant electrons
exist \cite{Eschrig2006}, and a strong correlation tendency towards phase separation that gives rise to fluctuating stripes remains elusive. 

\begin{figure*}[t]
\includegraphics[width=2\columnwidth]{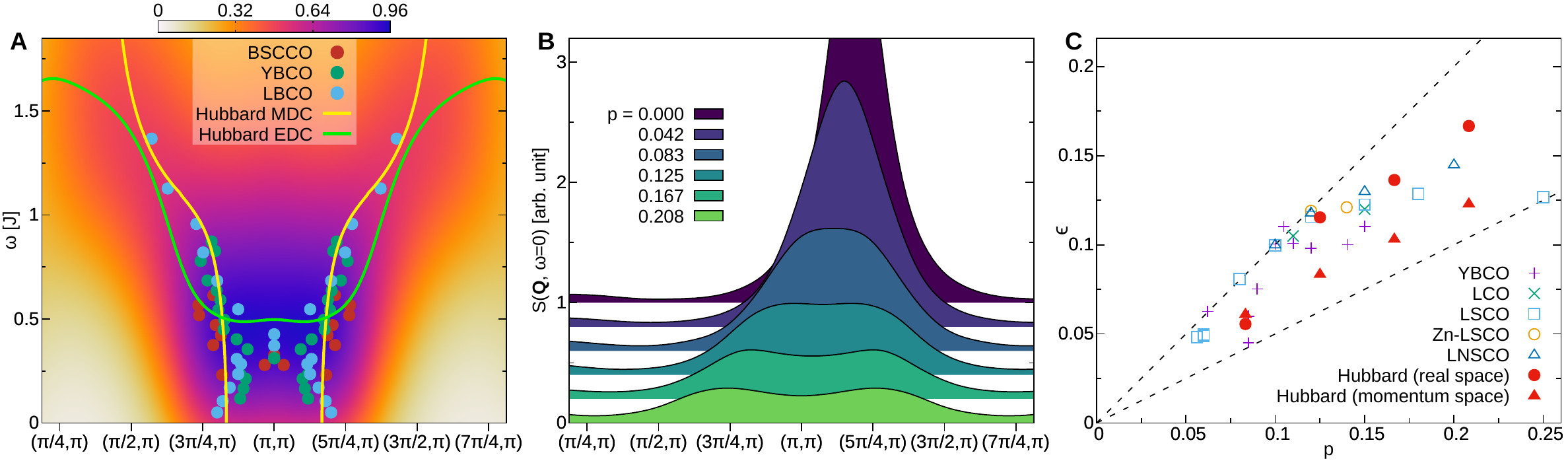}
\caption{{\bf Comparisons of the three-band Hubbard model against experimental results.\cite{Huang2016}} ({\bf A}) Dynamical spin structure factor $S(\bm{Q},\omega)$ (color plot) and EDC (energy distribution
curve) centers (green line) along $Q_y = \pi$ for $0.125$ hole doping. Yellow lines represent centers of double Gaussian fits to MDCs (momentum distribution curves). Dots show inelastic neutron scattering data\cite{Tranquada2007,Tranquada2004,Xu2009,Stock2005}. The Cu-O plane hole doping is $p \approx 0.16$ for optimally doped $\mathrm{Bi_2Sr_2CaCu_2O_{8+x}}$ [BSCCO \cite{Xu2009}], $p \approx 0.1$ for $\mathrm{YBa_2Cu_3O_{6.5}}$ [YBCO \cite{Stock2005}], and $p = 0.125$ for $\mathrm{La_{1.875}Ba_{0.125}CuO_4}$ [LBCO \cite{Tranquada2004}]. ({\bf B}) Plot of $S(\mathbf Q, \omega=0)$ for different dopings, with slight offsets for clarity. ({\bf C}) Low-energy spin incommensurability $\epsilon$ versus hole doping $p$. In units where $a = 1$, $\epsilon$ is the separation of low-energy peaks in $S(\mathbf Q, \omega)$ from $\mathbf Q = (\pi,\pi)$, divided by $2\pi$. Dashed lines show $\epsilon = p$ and $\epsilon = p/2$, corresponding to incommensurabilities of half-filled stripes \cite{White1998} and filled stripes \cite{Zaanen1989}, respectively. Data for YBCO \cite{Dai2001,Mook2002,Hinkov2008},
LCO \cite{Wells1997,Lee1999}, LSCO \cite{Yamada1998,Wakimoto1999,Wakimoto2000,Matsuda2000_1,Matsuda2000_2,Matsuda2002}, Zn-LSCO \cite{Kimura1999,Tranquada1999}, and LNSCO
\cite{Tranquada1995,Tranquada1996,Ichikawa2000} are plotted with estimates of incommensurability
from the DQMC data. The calculation of the real-space estimate is
based on half the inverse of antiphase domain wall periodicity. The momentum-space estimate is obtained by double Gaussian
fits to MDCs in (B). Neither method shows incommensurability
for $p = 0.042$.}
\label{f4}
\end{figure*}

Various numerical methods have been applied to the Hubbard model to study its low temperature and ground state properties. It is now widely accepted that different candidate ground states all lie close in energy \cite{Corboz2014,Zheng2017}, with small differences possibly associated with specific aspects of the numerical method used. Density matrix renormalization group, exact diagonalization/dynamical mean-field theory, constrained path auxiliary field Monte Carlo, infinite projected entangled-pair states, and density matrix embedding theory all find evidence for stripes \cite{White1998,White2003,Hager2005,Fleck2000,Chang2010,Corboz2014,Zheng2017}, having stronger amplitudes and longer correlation lengths than $d$-wave superconductivity. On the other hand, dynamical cluster approximation and cellular dynamical mean-field theory calculations suggest a finite temperature transition into a $d$-wave superconductor without stripe order in any portion of the phase diagram\cite{MaierRMP2005,Maier2005,Kancharla2008,Sordi2012,Gull2012,Gull2013}. 

Utilizing numerically exact DQMC to simulate interacting fermions on a square lattice at finite temperature gives an unbiased approach for investigating stripe order and its relationship to other possibly intertwined orders such as superconductivity. Despite the prevalence of the well-known fermion sign problem at low temperatures and dopings near half-filling, DQMC reveals a broad doping region of fluctuating stripe order that onsets roughly when temperatures are lowered through the magnetic exchange interaction $J/k_B$. Fluctuating stripe order is obtained in single-band and multi-band Hubbard models, different sizes and shapes of cluster as long as spin stripes can be accommodated, and for many combinations of parameters appropriate to those for the cuprates. We have found that the stripe period and correlation length is tuned largely by doping, with nearest-neighbor hopping $t'$ having also a strong effect.

The spin structure factor $S({\bf q})$ for copper and oxygen ligand orbitals from three-band DQMC on a $4\times 4$ square lattice Cu$_{16}$O$_{32}$ as a function of doping is plotted in Fig.~\ref{f1}.\cite{Kung2016} Copper $(\pi,\pi)$ antiferromagnetism dominates in the undoped system, decreasing strongly with doping - more so with hole doping than electron doping as observed in experiments.  The oxygen orbitals do not show signs of any particular spin order, again consistent with experiments. Similar behavior to that shown in Fig.~\ref{f1} is found for the single band Hubbard model, where oxygen degrees of freedom are integrated out.\cite{SBHAF1,SBHAF2} Therefore for spin-derived properties there are essentially little differences between single and multi-band Hubbard models.

Figure~\ref{f3} shows staggered spin correlation functions, defined as $\tilde S(i,j) = (-1)^{i+j} \langle S_z(i) S_z(j)\rangle$ with $S_z(i)=\frac{1}{2}(n_{i,\uparrow}-n_{i,\downarrow})$, at $1/8$ hole doping for various values of $t'$ in the single-band Hubbard model on $4\times16$ clusters with periodic boundary conditions, providing strong evidence for fluctuating spin stripe order as seen in experiments.\cite{Huang2018} 
DQMC simulations show that these short-range stripe correlations develop for temperatures below the antiferromagnetic exchange $J$ and do not depend strongly on the precise Hubbard $U$ apart from setting $J$. However, as shown in Fig.~\ref{f3},
the period of the stripe order can be altered by next nearest neighbor hopping $t'$, consistent with other numerical studies such as DMRG\cite{YY}. 
A similar dependence also manifests in the superconducting correlations, likewise tunable by $t'$\cite{XX}.

The dynamical structure factor for $4\times16$ 3-band Hubbard model Cu$_{64}$O$_{128}$ is shown in Fig.~\ref{f4}, demonstrating the development of magnetic incommensurabilities to yield split peaks away from the antiferromagnetic wavevector ${\bf Q = (\pi,\pi)}$ that increase monotonically with hole doping away from the antiferromagnetic insulating state at half-filling.\cite{Huang2016} A corresponding ``hourglass" spectra can be well reproduced from DQMC simulations, giving magnetic incommensurabilites that lie close to those experimentally observed in the cuprates. The data indicate that the stripe order lies between filled and half-filled stripes, which has been found to be crucial for superconducting correlations\cite{XX}.

\begin{figure}[b]
\includegraphics[width=.9\columnwidth]{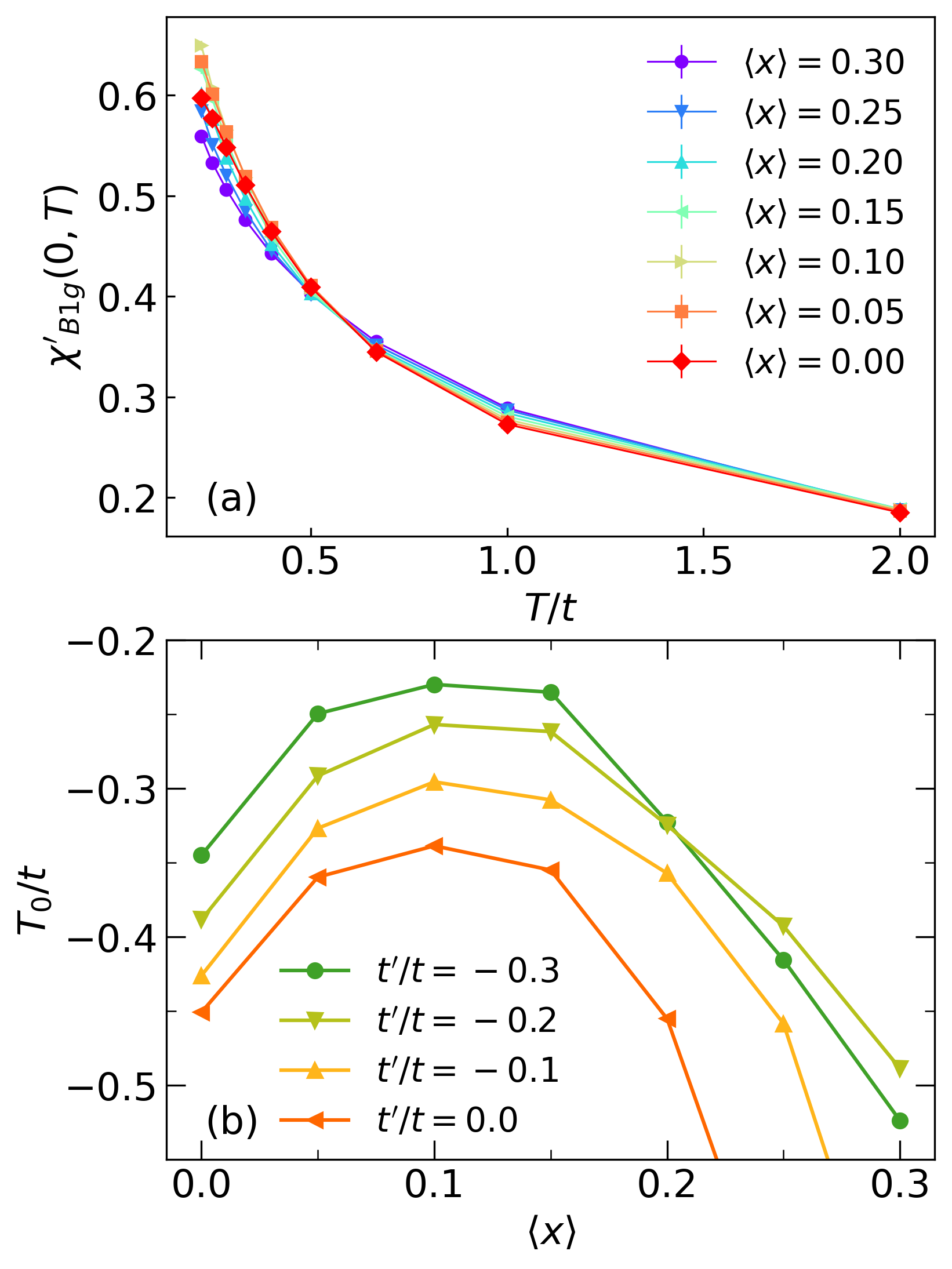}
\caption{(a) Nematic Raman $B_{1g}$ susceptibility calculated using DQMC as a function of temperatures for various doping levels $\langle x\rangle=1-n$, for $U/t=6$ and $t'/t=-0.25$. (b) Curie-Weiss temperatures of the $B{1g}$ susceptibility as a function of doping level for different next-nearest-neighbor hoppings.\cite{tianyi}}
\label{f5}
\end{figure}

Such a tendency to form fluctuating order at accessible temperatures may also be revealed in nematic states, ones in which the electron fluid breaks rotational symmetry while preserving translational invariance. Raman spectroscopy is a unique tool for probing nematic fluctuations and nematic ordering tendencies owing to its ability to symmetry resolve low energy excitations\cite{TPDRaman,tianyi}. The Raman nematic susceptibility $R_{B_{1g}}$ is defined as
\begin{equation}
    R_{B_{1g}} = \int_0^\beta d\tau \langle \rho_{B_{1g}}(\tau) \rho_{B_{1g}}(0) \rangle,
\end{equation}
where the effective scattering operator,
\begin{equation}
    \rho_{B_{1g}} = \frac{1}{2}\sum_{\boldsymbol{k}\sigma}(\cos k_x - \cos k_y) c^\dagger_\sigma(\boldsymbol{k})c_\sigma(\boldsymbol{k}),
\end{equation}
is a projected charge density operator with a $B_{1g}$-wave form factor that can capture broken $C_4$ rotational symmetry of the electrons. Fig.~\ref{f5} shows the $B_{1g}~(x^2-y^2)$ nematic susceptibility extracted from DQMC simulations for $8\times8$ clusters, which increases dramatically with decreasing temperature for all hole dopings.\cite{tianyi} The susceptibility reveals a non-monotonic doping dependence at fixed temperature. A peak in the nematic susceptibility is detected around $1/8$ doping, consistent with Raman experiments in the cuprates. The role of correlations on the nematic susceptibility can be assessed by comparing the full susceptibility to one approximated by evaluating the response function using renormalized Green's functions only, without the vertex corrections due to the Hubbard interaction. The latter "bubble" susceptibility does not show an anomalous doping dependence as is the case for the full response. In addition, at half-filling it decreases due to opening of the Mott gap and absence of two-magnon excitations that would be captured by the vertex corrections. This supports the notion of a strong tendency towards phase separation induced by strong local Hubbard interaction as a cause of stripe correlations.

\section{Pair field susceptibilities and pair density waves}

The interplay between superconductivity and these magnetic stripe and $B_{1g}$ nematic correlations has received a great deal of attention using DMRG. In doped quasi-one dimensional four-leg ladders a Luther-Emery phase has been identified with dual quasi-long-range charge density-wave order and superconductivity\cite{Fradkin2015,YY,PhysRevResearch.2.033073}. Filling stripes or decreasing $t'$ tends to decrease superconducting correlations at the expense of stripe correlations. In this next section we review DQMC calculations to examine superconducting correlations on top of the spin stripe and nematic correlations detected at finite temperatures.

\begin{figure}[t]
\includegraphics[width=0.95\columnwidth]{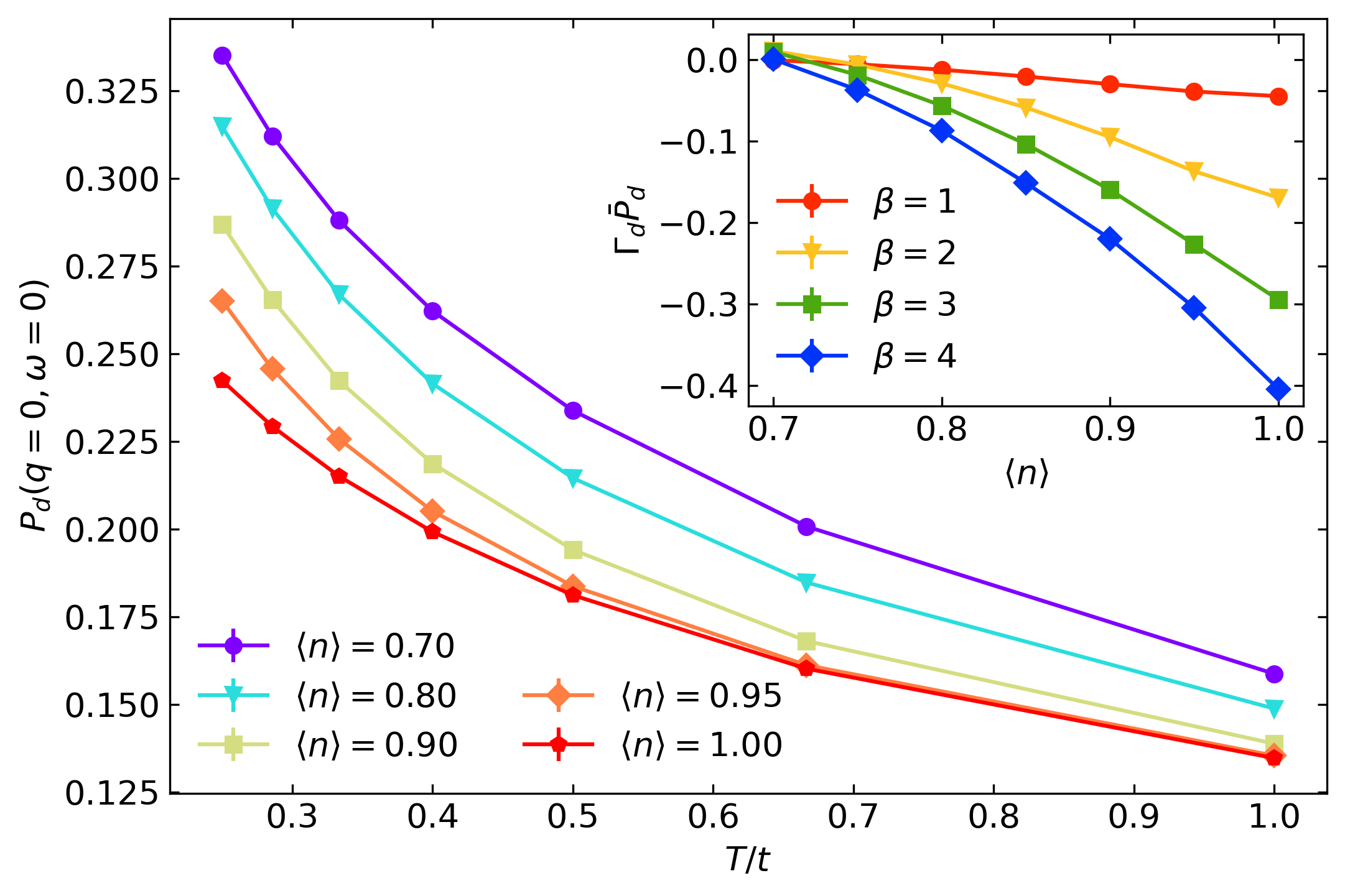}
\caption{$d-$wave pairing susceptibilities and vertex as a function of temperature for various dopings with parameters $U/t = 6$ and $t'/t = -0.25$ from DQMC simulations performed on $8\times8$ square clusters.}
\label{f6}
\end{figure}

Fig.~\ref{f6} plots DQMC results in-line with many prior studies, indicating that dominant $d_{x^2-y^2}$ pair-field susceptibilities rise with decreasing temperature. Here the pair-field susceptibility is defined as\cite{white1989attractive}
\begin{equation}
    P_d(\mathbf{q}=0, \omega=0) = \frac{1}{N} \int_0^\beta d\tau \langle \Delta_d(\tau)\Delta_d^\dagger(0) \rangle,
\end{equation}
\begin{equation}
    \Delta_d^\dagger = \sum_{\mathbf{k}}(\cos{k_x}-\cos{k_y})c_{\mathbf{k},\uparrow}^\dagger c_{\mathbf{k},\downarrow}^\dagger.
\end{equation}

The non-interacting system has a divergent pair-field susceptibility at low temperatures due to the nesting of the Fermi surface and van Hove singularity in the density of states, so an infinitesimal attractive interaction can drive it into a superconducting state; on the contrary, the repulsive Hubbard U suppresses the pair-field susceptibility \cite{hirsch1985two, hirsch1988pairing}. 

Differences between the interacting and non-interacting susceptibilities may be revealed considering the superconducting vertex
\begin{equation}
    \Gamma_d = \frac{1}{P_d} - \frac{1}{\bar{P}_d},
\end{equation}
where $\bar{P}_d$ is the uncorrelated pair-field susceptibility. 
The inset to Fig.~\ref{f6} displays the superconducting vertex, revealing that pairing correlations are largest for these temperatures when approaching half-filling, developing in-step with the establishment of strong anti-ferromagnetic correlations.  

While we cannot access temperatures low enough to determine whether a true superconducting instability occurs at lower temperature, we can estimate how close one might be to a true transition by examining the dynamical pair-field susceptibility, obtained via analytical continuation of the imaginary time pair field correlators\cite{jarrell1996bayesian}:
\begin{equation}
    P_d(\mathbf{q} = 0, i\omega_n) = \int_0^\beta d\tau e^{i\omega_n\tau} \frac{1}{N} \langle \Delta_d(\tau)\Delta_d^\dagger(0) \rangle,
\end{equation}
\begin{equation}
    P_d(\mathbf{q} = 0, \omega) = P_d(\mathbf{q} = 0, i\omega_n \rightarrow \omega + i\delta).
\end{equation}

\begin{figure}[t]
\includegraphics[width=0.95\columnwidth]{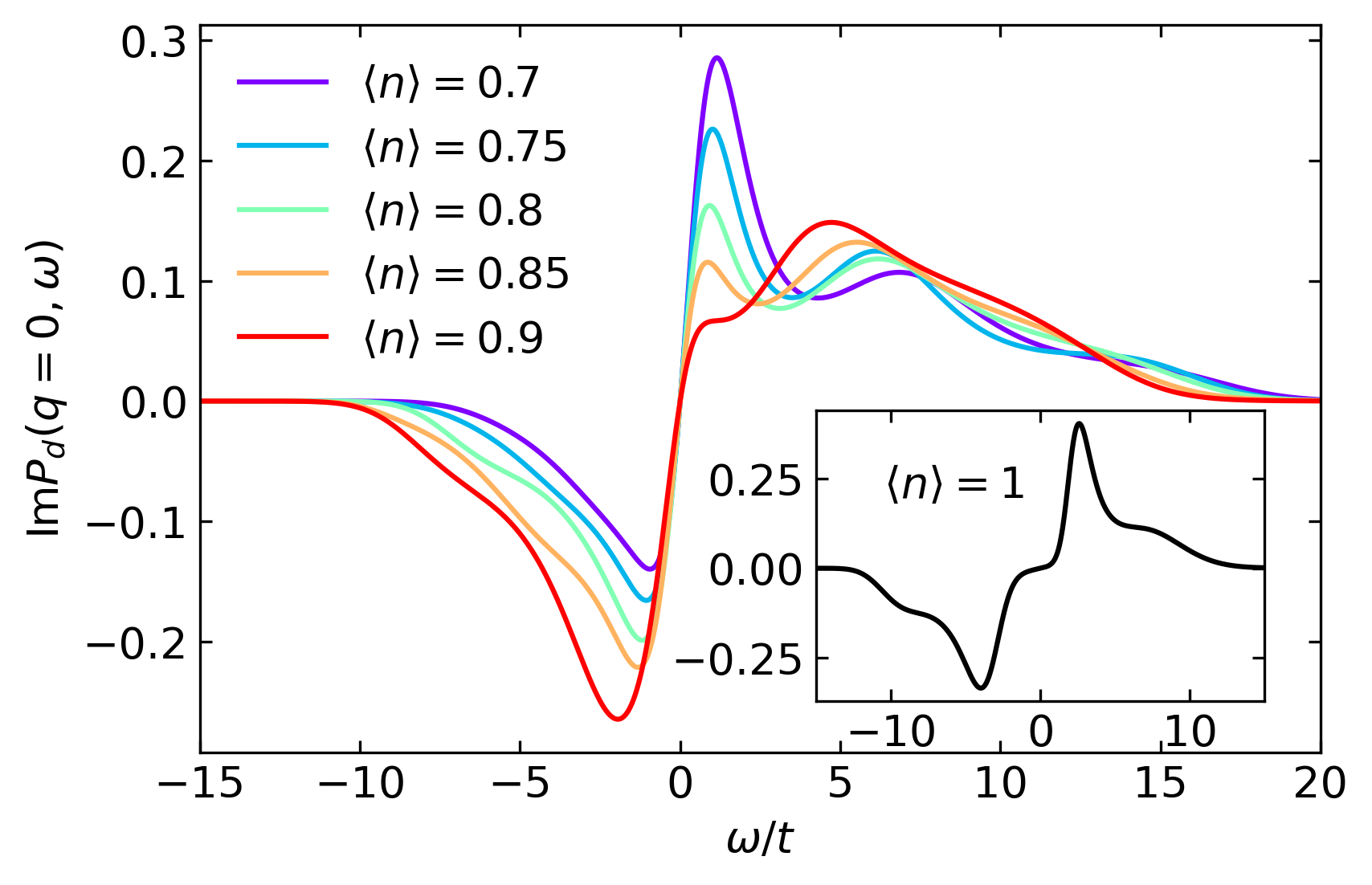}
\caption{Dynamical $d-$wave pairing susceptibility for various dopings at $\beta=4.5/t$, from DQMC simulations on $8\times8$ square clusters for $t'/t=-0.25$}.
\label{f7}
\end{figure}

Fig.~\ref{f7} plots the dynamical pair-field susceptibility on an $8\times8$ cluster, revealing a peak at frequencies $\sim U$ and a low energy peak that rises and moves to lower frequencies with progressive doping away from half filling. In principle the frequency at which the dynamical pair-field susceptibility is peaked is a measure of the ``mass" of superconducting pair fluctuations. Apart from a strong peak at the Hubbard $U$ at half-filling, the low frequency peak that emerge from half-filling and rises at lower energies, growing in intensity with increasing doping, can be associated with a distance from a superconducting instability (when the mass vanishes). Such pair-field dynamics may be measured either in pair-tunneling experiments\cite{scalapino} or two-photon pair photoemission\cite{pair}, although as far as we are aware to date no such experiments have been performed on the cuprates.

As can be seen from both the static and dynamic information, finite temperature spin stripe/nematic fluctuations follow similar tendencies as these superconducting fluctuations, and similar to results at $T=0$ from DMRG.\cite{XX} There is a smooth doping evolution, with no particular emphasis at commensurate fillings such as $1/8$. 

\begin{figure}[t]
\includegraphics[width=\columnwidth]{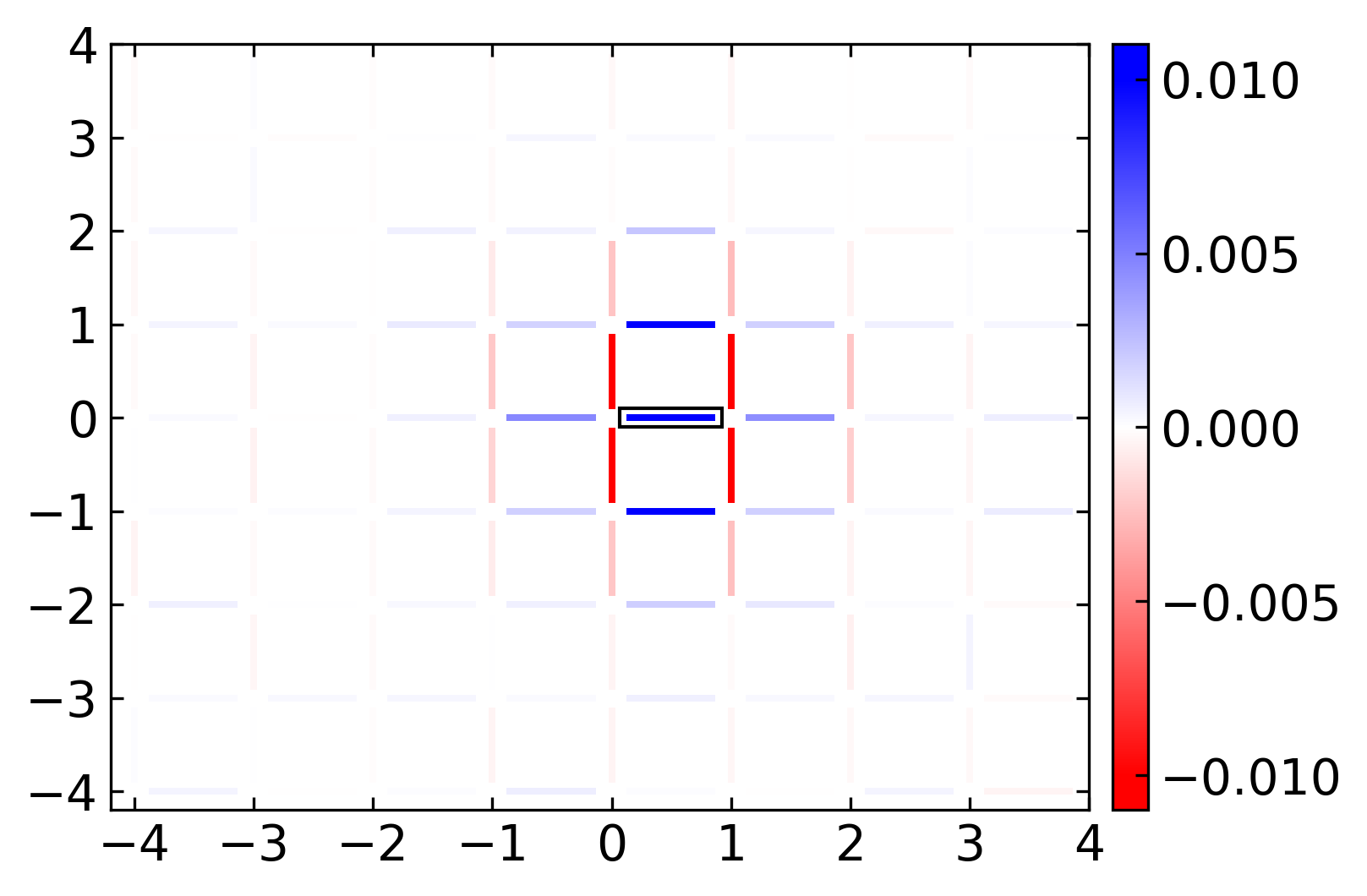}
\caption{The singlet-pair correlation function in real space for $U/t=6$, $\beta=4.5/t$, $t'/t=-0.25$ and $\langle n \rangle = 0.85$. The black box highlights the reference bond. A local $d-$wave pattern is observed.}
\label{f8}
\end{figure}

Contradictory evidence from scanning tunneling measurements\cite{edkins2019magnetic} and x-ray scattering measurements\cite{stegen2013evolution} in cuprates has raised the issue of the presence of a possible pair-density wave (PDW) - a superconducting paired state having finite center-of-mass momentum \cite{agterberg2020physics}. The intertwined nature of the PDW states poses difficulties on numerically identifying it in models for strongly correlated systems. In the $t\ensuremath{-}J$ model, though DMRG studies have yet to observe a PDW \cite{PhysRevB.95.155116}, iPEPS calculations indicate that its ground state energy may be very close to the competing striped Luther-Emery phases \cite{corboz2014competing}. Only one-dimensional Kondo-Heisenberg chain \cite{PhysRevLett.105.146403} and Holstein-Hubbard model on triangular cylinders \cite{huang2021pair} have been shown to have a quasi-long-range PDW order from DMRG calculations.

In Fig.~\ref{f8} we plot real space singlet-pair correlations on the same cluster, defined as
\begin{equation}
    S_{\alpha,\alpha'}(\mathbf{r}) = \frac{1}{N}\int_0^\beta d\tau \sum_i \langle \Delta_{\alpha}(\tau, \mathbf{r}_i + \mathbf{r}) \Delta_{\alpha'}^\dagger(0, {\mathbf{r}_i}) \rangle,
\end{equation}
\begin{equation}
    \Delta_{\alpha}^\dagger(\mathbf{r}_i) = \frac{1}{\sqrt{2}} \left(c_{\mathbf{r}_i,\uparrow}^\dagger c_{\mathbf{r}_i+\alpha,\downarrow}^\dagger - c_{\mathbf{r}_i,\downarrow}^\dagger c_{\mathbf{r}_i+\alpha,\uparrow}^\dagger\right),
\end{equation}
\begin{equation}
\alpha = \hat{x} \text{ or } \hat{y}.
\end{equation}
for $15\%$ doping in DQMC, where clear short-range $d_{x^2-y^2}$ patterns are observed, decaying after a few lattice spacings, with faint signatures remaining near the sizes of our cluster (here $8\times 8$, with no qualitative changes for larger clusters).

\begin{figure}
\includegraphics[width=\columnwidth]{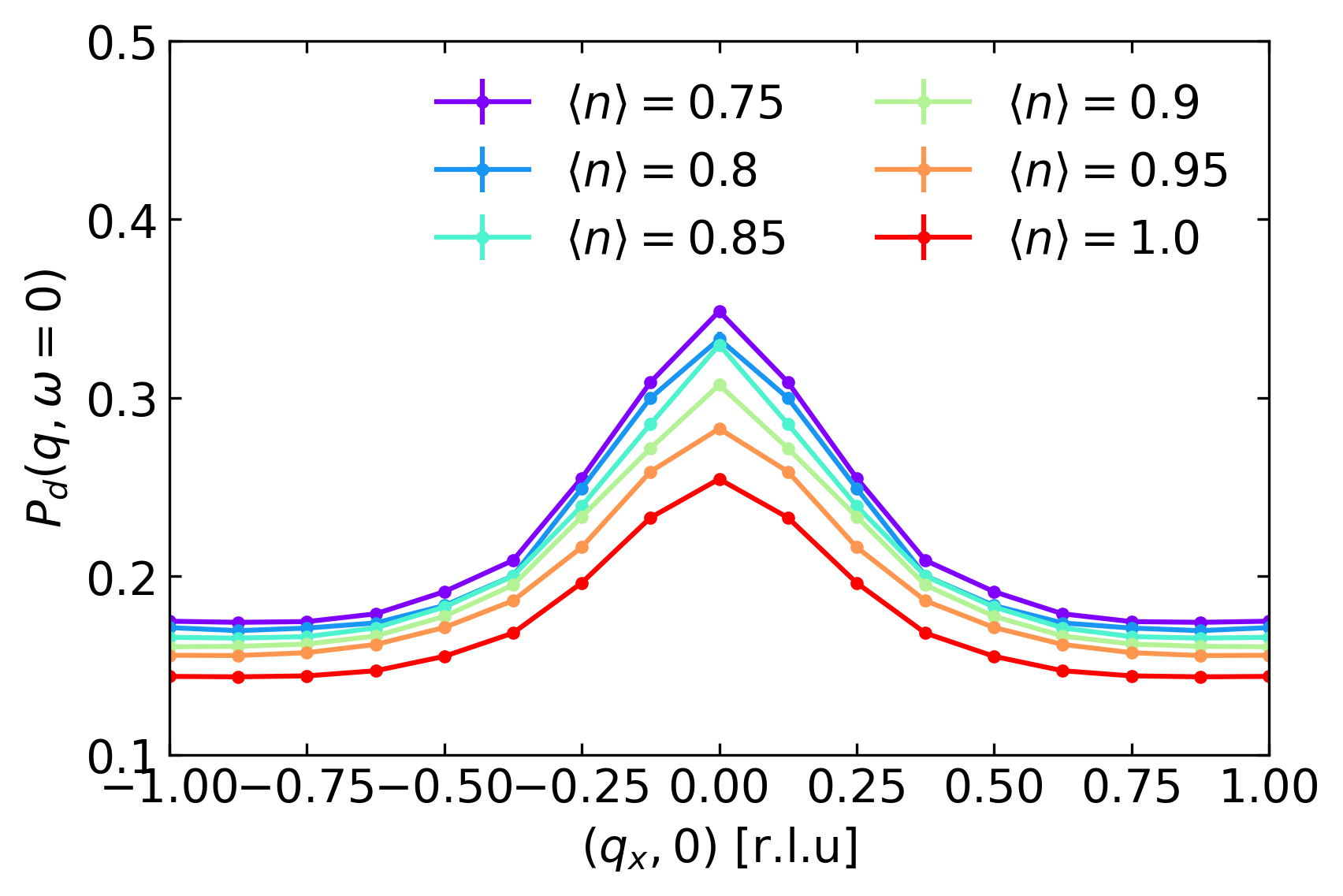}
\caption{Fourier transformed $d-$wave pair-field susceptibilities for different dopings for $U/t=6$, $\beta=4.5/t$, and $t'/t=-0.25$, plotted along the Brillouin zone boundary. The pairing peaks at $(0, 0)$.}
\label{f9}
\end{figure}

Fourier transforming this pattern in Fig.~\ref{f9} shows that the pairs have a dominant zero center-of-mass momentum, with no indication of a PDW at any other momentum for these temperatures. Again we see congruent behavior at zero temperature on quasi-one dimensional ladders using DMRG and at finite temperatures on two-dimensional clusters using DQMC: PDW correlations are much weaker than stripe and zero center-of-mass superconducting correlations.

\begin{figure}[b]
\includegraphics[width=\columnwidth]{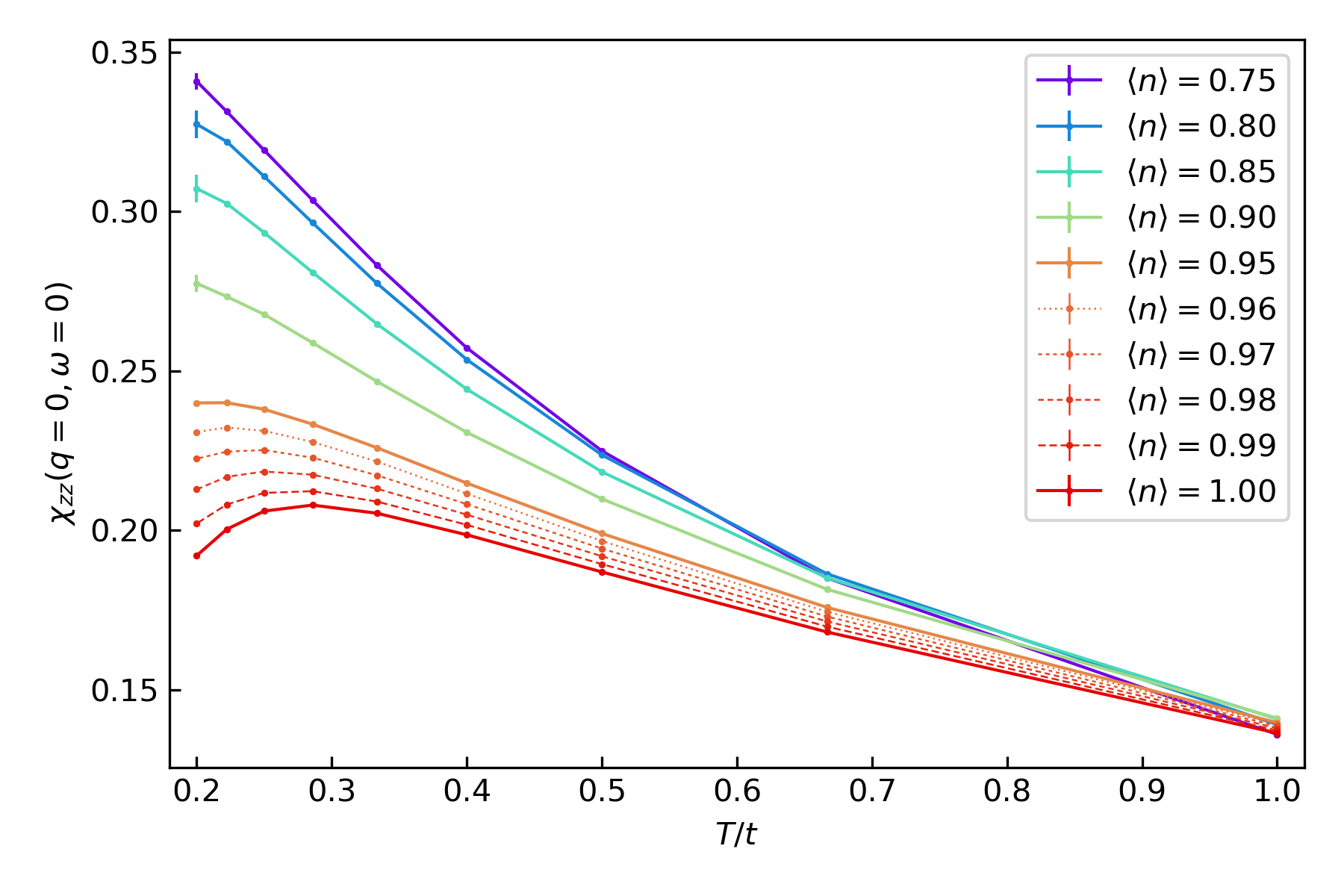}
\caption{Spin (Pauli) susceptibility, or Knight shift, as a function of temperature for different dopings, with parameters $U/t=6$ and $t'/t=-0.25$. 
Simulation cluster size is $8 \times 8$.
The Knight shift shows a peak at low temperatures, identified as $T^*$, at half-filling that moves to lower temperatures for increasing doping. Due to the fermion sign problem, it cannot be tracked from DQMC for dopings above $5\%$. The peak temperature $T^*$ is not correlated with the development of superconducting or nematic/stripe correlations.}
\label{f10}
\end{figure}

\begin{figure*}[h]
\centering
\includegraphics[width=1.6\columnwidth]{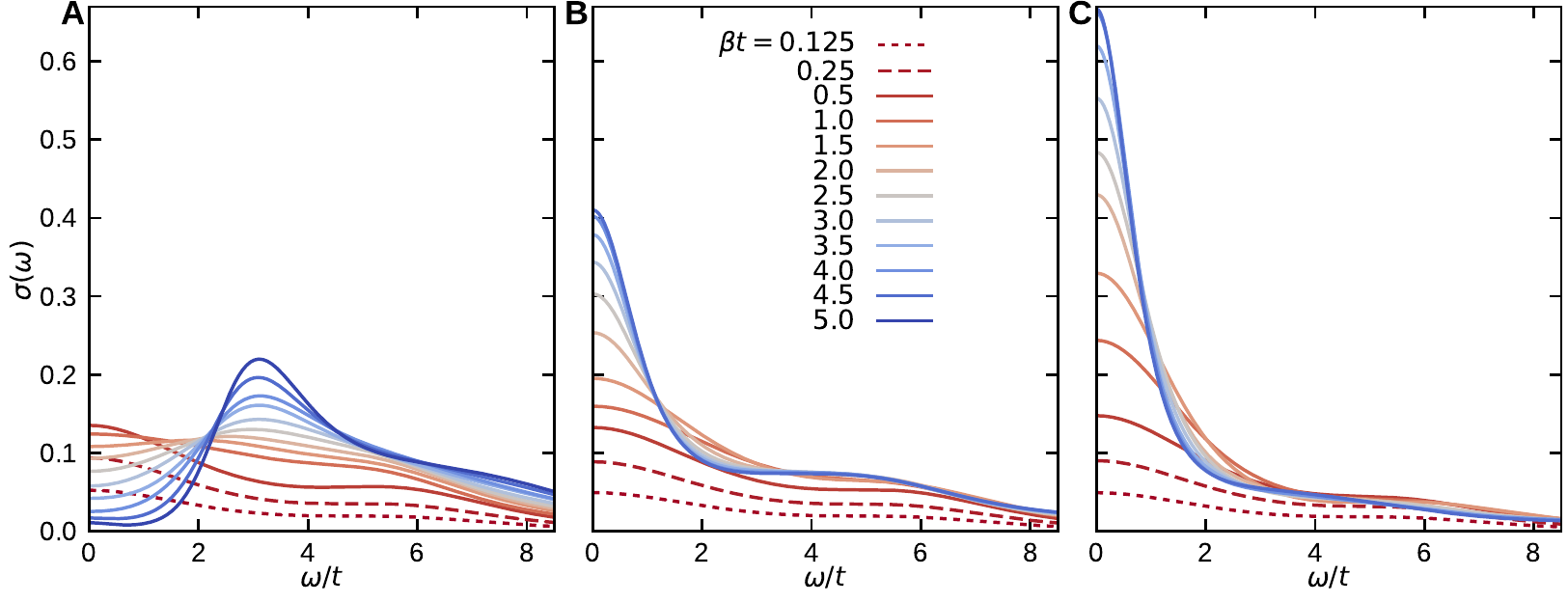}
\caption{Optical conductivity obtained through DQMC
and maximum entropy analytic continuation for the Hubbard model with parameters 
$U/t = 6, t'/t = -0.25$.
Shown are the data for various temperatures $T = 1/\beta$, where Boltzmann’s constant $k_B = 1$. Hole
doping level is (\textbf{A}) $p = 0.0$, (\textbf{B}) $0.1$, and (\textbf{C}) $0.2$. Simulation cluster size is $8 \times 8$.\cite{Huang2019}}
\label{f11}
\end{figure*}

Finally, while we note that spin stripe, nematic, and superconducting fluctuations are predominant over a wide range of doping, the Knight shift shows important differences. As one of the key indicators of a ``pseudogap" phase in cuprates, defining the crossover temperature $T^*$ from its peak, this has been found to decrease with increasing doping.
The Knight shift $\chi$ is calculated from DQMC from the local static magnetic susceptibility via
$$
\chi = \frac{1}{N}\sum_i \int_0^{\beta} d\tau \langle S_z(i,\tau) S_z(i,0)\rangle.
$$
Our DQMC results shown in Fig.~\ref{f10} reveal a peak in the Knight shift at low temperatures that decreases with increasing doping, at least over the limited accessible range near half-filling where the peak can be tracked due to the fermion sign problem.  Importantly, we note that the onsets of stripe, nematic, and superconducting fluctuations do not seem to be correlated with $T^*$.  It is very much of an open question how the pseudogap region - a large region over the phase diagram of the cuprates - affects ordering tendencies and transport properties. We review DQMC calculations for transport in the next section.

\section{Transport in the ``strange metal" phase}

Strongly correlated materials are not only renowned for their rich phase diagrams containing intertwined orders but also the tendency to be "bad" metals\cite{Fradkin2015,Keimer2015} at higher temperatures. A key towards understanding the emergent orders at lower temperatures lies largely in the anomalous properties of the high temperature disordered phase: DC resistivity in the normal state that exceeds the Mott-Ioffe-Regel (MIR) criterion with no sign of a crossover or saturation, signaling the absence of well-defined quasiparticles\cite{Gunnarsson2003,Hussey2004}. In a number of these systems, the resistivity generally can be characterized by a linear-in-temperature dependence to the highest accessible temperatures. 
These features are incompatible with conventional Fermi liquid theory and pose a fundamental challenge to the understanding of strongly correlated materials. Early on, it was recognized that the maximal transition temperatures in hole-doped cuprates occur in a region of doping with the highest prevalence of $T$-linear resistance.  This is suggestive of at least a deeper connection between the strange metal state and the origin of pairing in unconventional superconducting materials.
This motivates a more detailed study of resistivity and transport properties of the strongly correlated Hubbard model.

\begin{figure}[b]
\centering
\includegraphics[width=\columnwidth]{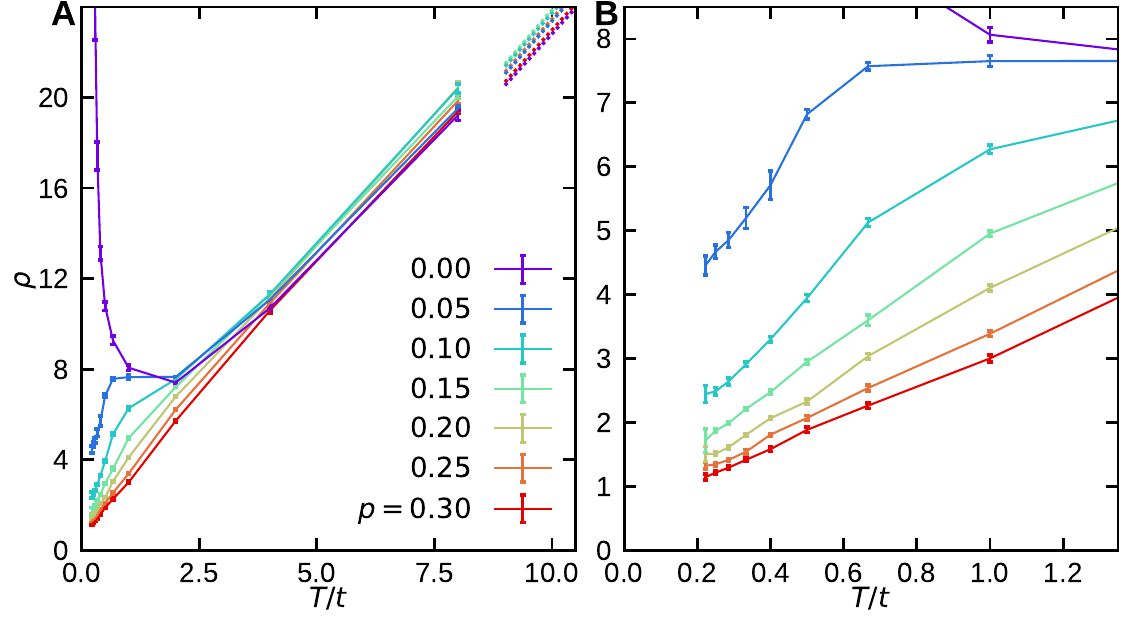}
\caption{(\textbf{A}) DC resistivity as a function of
temperature and hole doping, obtained from analytically continued optical conductivity as shown
in Fig.~\ref{f11}, from DQMC simulations on $8 \times 8$ clusters. The resistivity $\rho$ is plotted in units of $\hbar/e^2$. Solid lines through DQMC data points are guides to the eye. Dotted lines are results
from moments expansions up to $18$th order in the high-temperature limit\cite{Huang2019}. (\textbf{B}) Close-up view
of the lowest-temperature data of (\textbf{A}). Error bars represent random sampling errors, determined
by bootstrap re-sampling.\cite{Huang2019}}
\label{f12}
\end{figure}

\begin{figure*}[t]
\includegraphics{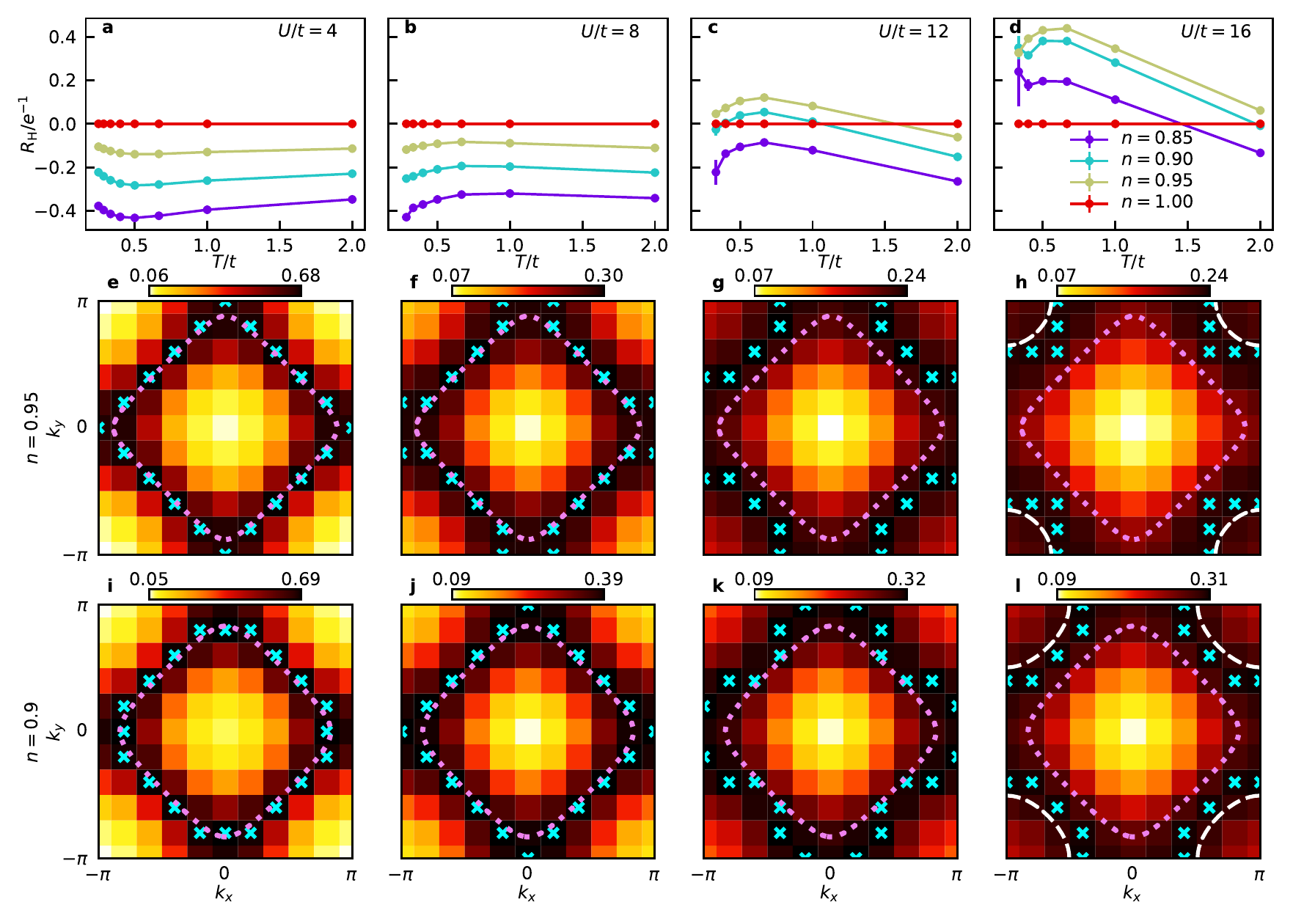}
\caption{Temperature and doping dependence of the Hall coefficient $R_H$ for $U/t=4-16~$(\textbf{a}-\textbf{d}, simulation cluster size $N=8\times8$)\cite{Wang2020,assa,assa2} without next nearest neighbour hoping $t'=0$, connecting to Fermi surface positions approximated by the the peak positions (cyan crosses) of the imaginary time green function $G(\mathbf{k},\tau=\beta/2)\beta$ in \textbf{e}-\textbf{l} (Temperature $T=0.5t$, simulation cluster size $10\times 10$).
Panels in the same column are for the same interaction.
\cite{footnoteaboutdtau} 
Pink dotted lines are the Fermi surface for the non-interacting tight binding model for the corresponding filling level.
White dashed lines are the Fermi surface predicted by the Hubbard-I approximation.\cite{Wang2021}}
\label{f13}
\end{figure*}

To extract the optical conductivity from the imaginary time current-current correlator measured using DQMC, we employ the standard method of maximum entropy analytic continuation\cite{Huang2019,Jarrell1996,Gunnarsson2010}. In Fig.~\ref{f11} we show temperature dependence of AC optical conductivity at (a) $p = \langle x\rangle = 1-n = 0$ (half-filling), (b) $p=0.1$, and (c) $p=0.2$. 
At half-filling, 
insulating behavior sets in below roughly $T \sim t$, where the DC conductivity begins to decreases at low frequencies due to the formation of a Mott gap. This contrasts with the more metallic behavior shown in the doped cases (Fig.~\ref{f11}B and C), where the conductivity at low frequencies monotonically increases with decreasing temperature leading to the formation of a Drude-like peak. At these filling levels, a secondary peak appears that can be associated with transitions between the Hubbard bands, but below $T \sim t$ relatively little spectral weight is transferred either to or from this peak as the temperature decreases.

High temperatures, above $T \sim 0.5t$, display a distinct metallic behavior for all doping, including half-filling.
This regime is marked by broad peaks at both low ($\omega = 0$) and high ($\omega \sim U$) frequency, with spectral weight that scales together with variations in temperature. In contrast to the metallic behavior at higher doping and low temperature, the width of this zero-frequency peak and overall frequency profile of the optical conductivity does not evolve with temperature.

The 
behavior of the optical conductivity at low frequency motivates further investigation of DC transport properties in the model. Figure~\ref{f12} shows the resistivity of the Hubbard model in natural units ($\hbar/e^2$) plotted versus temperature for a range of hole doping levels from $p=0$ (half-filling) to $p=0.30$. As a guide, the MIR limit is of order unity in natural units. Here, there is no evident saturation related to the MIR criterion in the data; and in fact, the resistivity at low doping still exceeds the MIR limit at the lowest accessible temperatures.

There exists a clear distinction in the behavior for temperatures below and above $T \sim 1.5 t$, which  marks something of an onset in insulating behavior at half-filling. 
In Fig.~\ref{f12}, we see that the doping dependence of DC resistivity also has a low and high temperature contrast, similar to the behavior observed in Fig.~\ref{f11}. While there is a considerable doping range of $T$-linear resistivity at both high and low temperatures, only low temperatures show a significant doping dependence, whereas the behavior at high temperatures is relatively doping independent. 
Two observations -- $T$-linear resistivity and a lack of saturation at the MIR limit -- indicate that a significant portion of the phase diagram for the Hubbard model displays strange metallic transport, at least in the regime of intermediate to strong Hubbard $U$. It is an interesting question to ask at what point in doping does one see a crossover to Fermi-liquid like transport, which would require further analysis at higher dopings. We might expect that finite size effects might then become more relevant as the energy gap between adjacent momentum points will play a dominant role in setting thermal transport properties. This is an interesting case for further study.

To further test the properties of transport without a coherent quasiparticle picture as in weak coupling, we have explored the Hall coefficient via DQMC simulations of the Hubbard model. There are several ways that it can be calculated, either in the limit of vanishing magnetic fields\cite{assa,assa2,3currentderivation,3currentderivation2,3currentderivation3} or by direct inclusion of the magnetic field into the DQMC algorithm. We have performed a careful study\cite{Wang2021} showing similar behaviors at low fields can be obtained by these methods. Fig.~\ref{f13} (top panel) displays the Hall coefficient obtained from DQMC\cite{Wang2020,assa,assa2} for different values of the Hubbard $U$ and dopings as a function of temperature. In contrast with weak coupling Boltzmann transport, the Hall coefficient develops a strong temperature dependence for stronger $U$, rising as temperature is lowered - a behavior similar to that observed in the cuprates.\cite{Hwang1994}

Moreover, while the Hall coefficient is negative for weaker $U$, reflecting electron Fermi surfaces centered around $\Gamma$, the Hall coefficient can change sign for stronger $U$ and become positive, indicating a Fermi surface reconstruction as temperature is lowered to a more hole-like behavior. DQMC simulations confirm a linkage between the closing of the Fermi surface and the sign of the Hall coefficient at all temperatures and dopings even in the presence of a strong Hubbard $U$ which disfavors coherent quasiparticles and gives resistivities well past the MIR limit.

As $U$ becomes stronger,
the Fermi surface changes from electron-like (centered at $\Gamma$
point) to hole-like (centered at $M$ point) (Fig.~\ref{f13}\textbf{e} to \ref{f13}\textbf{h} for filling $n=0.95$ or \ref{f13}\textbf{i} to \ref{f13}\textbf{l} for filling $n=0.9$).
As the Fermi surface topology changes, the sign of $R_H$ changes
from negative to positive for these fillings(\ref{f13}\textbf{a} to \ref{f13}\textbf{d}), while $R_H$ always vanishes for half filling due to the particle hole symmetry.
Therefore a clear connection continues to exist between
$R_H$ and the Fermi surface topology, even without a well-defined Fermi
surface or well-formed quasiparticles.

\section{Perspective and outlook}
 Signatures of fluctuating order, which have been observed at relatively high temperatures in our model calculations, represent a strong piece of evidence that stripes generally should have strength enough to affect all electronic properties in the phase diagram. Our state-of-the-art numerical calculations have shown that even in fluctuating form stripes possess characteristic antiphase periodicity, while being robust to changes in cluster size and boundary conditions. They also are robust to changes in model parameters, which alter details, but do not affect their existence over a wide portion of the phase diagram.  Such fluctuating structures also may have a bearing on the true ground state of the Hubbard model and controversies that have arisen between previous numerical investigations\cite{Corboz2014}.
 However, to go beyond comparisons based solely on static properties requires an accurate numerical benchmark of dynamical properties\cite{Leblanc2015}.

In remarkable analogy to experimental phase diagrams\cite{YY,Zheng2017,Huang2018} ground state calculations of the Hubbard model have revealed intertwined orders, but important questions remain concerning how such order may emerge from the normal state. Controlled approaches, such as our DQMC calculations, suffer from a severe sign problem and currently are unable to directly access the relevant phases. Does superconductivity follow directly from the strange metal in the Hubbard model? Do coherent quasiparticles emerge between the strange metal and the ground state? These remain intriguing, open questions. We may find some answers to these questions by extending our measurements of dynamical quantities, including resistivity, through the development of new numerical techniques or via improved quantum simulations, which offer hope of overcoming some of the more severe limitations of current methods.

Strange metallicity in the Hubbard model at low temperatures, at least when compared to the other energy scales in the problem, provides promising support that the fundamental physics of real correlated materials may be approached through the study of simplifying, model Hamiltonians, which focus on relevant low energy degrees of freedom. In that regard, the numerical results presented here provide an important benchmark for theoretical descriptions of strange metals and approximate approaches to the solution of the Hubbard model\cite{Prushke1995,Bergeron2011,Deng2013,Xu2013}. Recent developments have even seen the measurement of transport properties in the Hubbard model using properly developed cold atoms experiments\cite{Xu2016,Brown2018,Nichols2018}. While those findings are broadly similar to our results, both finite temperature numerical approaches and cold atoms experiments will need to overcome significant obstacles to study normal state phenomena down to temperatures proximate to the superconducting temperature or those of other emergent phases.

\begin{acknowledgment}
This work was supported by the US Department of Energy, Office of Basic Energy Sciences, Materials Sciences and Engineering Division, under Contract No. DE-AC02-76SF00515. EWH was supported by the Gordon and Betty Moore Foundation EPiQS Initiative through the grants GBMF 4305 and GBMF 8691.
\end{acknowledgment}

\bibliographystyle{jpsj}
\bibliography{dqmcrefs}

\end{document}